\documentclass[runningheads]{llncs}

\usepackage[T1]{fontenc}
\usepackage{graphicx}
\usepackage{graphicx}

\usepackage{mathtools}
\usepackage[utf8]{inputenc} 
\usepackage[T1]{fontenc}    
\usepackage{url}            
\usepackage{booktabs}       
\usepackage{amsfonts}       
\usepackage{nicefrac}       
\usepackage{microtype}      
\usepackage{color}
\usepackage{bm}
\usepackage{bbm} 
\usepackage{bigstrut}
\usepackage{multirow}
\usepackage{amsmath}
\usepackage{colortbl}
\usepackage{mathrsfs}   
\usepackage{footnote}
\usepackage{setspace}
\usepackage{algorithmic}
\usepackage[linesnumbered,ruled]{algorithm2e}
\usepackage{array}
 \usepackage[T1]{fontenc}
 \usepackage{lmodern}
 \usepackage[misc]{ifsym}

\begin{document}

\title{Multi-task Learning of Histology and Molecular Markers for Classifying Diffuse Glioma}

\titlerunning{Multi-task learning for Classifying Diffuse Glioma}

\author{Xiaofei Wang\inst{1} \and
Stephen Price\inst{1}\and
Chao Li\inst{1,2,3,4}}

\institute{Department of Clinical Neurosciences,
University of Cambridge, UK \and
Department of Applied Mathematics and Theoretical Physics, University of Cambridge, UK \and
School of Science and Engineering, University of Dundee, UK \and
School of Medicine, University of Dundee, UK
}

\maketitle              
\begin{abstract}
Most recently, the pathology diagnosis of cancer is shifting to integrating molecular makers with histology features. It is a urgent need for digital pathology methods to effectively integrate molecular markers with histology, which could lead to more accurate diagnosis in the real world scenarios. This paper presents a first attempt to jointly predict molecular markers and histology features and model their interactions for classifying diffuse glioma bases on whole slide images. Specifically, we propose a hierarchical multi-task multi-instance learning framework to jointly predict histology and molecular markers. Moreover, we propose a co-occurrence probability-based label correction graph network to model the co-occurrence of molecular markers. Lastly, we design an inter-omic interaction strategy with the dynamical confidence constraint loss to model the interactions of histology and molecular markers. Our experiments show that our method outperforms other state-of-the-art methods in classifying diffuse glioma,as well as related histology and molecular markers on a multi-institutional dataset.

\keywords{Diffuse Glioma  \and Digital Pathology \and Multi-task learning \and Muti-label Classification.}
\end{abstract}

\section{Introduction}

Diffuse glioma is the most common and aggressive primary brain tumors in adults, accounting for more deaths than any other type \cite{wwwcancer}. Pathology diagnosis is the gold standard for diffuse glioma but is usually time-consuming and highly depends on the expertise of senior pathologists \cite{liang2019clinical}. Hence,  automatic algorithms based on histology whole slide images (WSIs) \cite{lu2021data}, namely digital pathology, promise to offer rapid diagnosis and aid precise treatment.

 Recently, deep learning has achieved success in diagnosing various tumors \cite{campanella2019clinical,yang2021deep}. Most methods are mainly predicting histology based on WSI, less concerning molecular markers. However, the paradigm of pathological diagnosis of glioma has shifted to molecular pathology, reflected by the 2021 WHO Classification of Tumors of the Central Nervous System \cite{louis20212021}. The role of key molecular markers, i.e, isocitrate dehydrogenas (IDH) mutations, co-deletion of chromosome 1p/19q and homozygous deletion (HOMDEL) of cyclin-dependent kinase inhibitor 2A/B (CDKN), have been highlighted as major diagnostic markers for glioma, while histology features that are traditionally emphasized are now considered as reference, although still relevant in many cases. For instance, in the new pathology scheme, glioblastoma is increasingly diagnosed according to IDH mutations, while previously its diagnosis mostly relies on histology features, including necrosis and microvascular proliferation (NMP). 
 \footnote{Similar changes of the diagnostic protocol can also be found in endometrial cancer \cite{imboden2021implementation}, renal neoplasia \cite{trpkov2021new}, thyroid carcinomas \cite{volante2021molecular}, etc.}.

 However, the primary approaches to assess molecular markers include gene sequencing and immuno-staining, which are time-consuming and expensive than histology assessment. As histology features are closely associated with molecular alterations, algorithm predicting molecular markers based on histology WSIs is feasible and have clinical significance. Moreover, under the new paradigm of integrating molecular markers with histological features into tumor classification, it is helpful to model the interaction of histology and molecular makers for a more accurate diagnosis. Therefore, there is an urgent need for developing novel digital pathology methods based on WSI to predict molecular markers and histology jointly and modeling their interactions for final tumor classification, which could be valuable for the clinically relevant diagnosis of diffuse glioma.

This paper proposes a deep learning model (DeepMO-Glioma) for glioma classification based on WSIs, aiming to reflect the molecular pathology paradigm. Previous methods are proposed to integrate histology and genomics for tumour diagnosis \cite{jiang2021predicting,xing2022discrepancy,chen2020pathomic}. For instance, Chen \textit{et al.}  \cite{chen2020pathomic} proposed a multimodal fusion strategy to integrate WSIs and genomics for survival prediction. Xing \textit{et al.} \cite{xing2022discrepancy} devised a self-normalizing network to encode genomics. Nevertheless, most existing approaches of tumor classification only treat molecular markers as additional input, incapable to simultaneously predict the status of molecular markers, thus clinically less relevant under the current clinical diagnosis scheme. 
To jointly predict histology and molecular markers following clinical diagnostic pathway, we propose a novel hierarchical multi-task multi-instance learning (HMT-MIL) framework based on vision transformer \cite{dosovitskiy2020image}, with two partially weight-sharing parts to jointly predict molecular markers and histology.

Moreover, multiple molecular markers are needed for classifying cancers, due to complex tumor biology. To reflect real-world clinical scenarios, we formulate predicting multiple molecular markers as a multi-label classification (MLC) task. Previous MLC methods have successfully modeled the correlation among labels \cite{li2022multi,yazici2020orderless}. For example, Yazici \textit{et al.} \cite{yazici2020orderless} proposed an orderless recurrent method, while Li \textit{et al.} designed a label attention transformer network with graph embedding. In medical domain, Zhang \textit{et al.} \cite{zhang2023triplet} devised a dual-pool contrastive learning for classifying fundus and X-ray images. Despite success, when applied to predicting multiple molecular markers, most existing methods may ignore the co-occurrence of molecular markers, which have intrinsic associations \cite{yip2012concurrent}. Hence, we propose a co-occurrence probability-based, label-correlation graph (CPLC-Graph) network to model the co-occurrence of molecular markers, i.e, intra-omic relationship. 

Lastly, we focus on modeling the interaction between molecular markers and histology. Specifically, we devise a novel inter-omic interaction strategy to model the interaction between the predictions of molecular markers and histology, e.g., IDH mutation and NMP, both of which are relevant in diagnosing glioblastoma. Particularly, we design a dynamical confidence constraint (DCC) loss that constrains the model to focus on similar areas of WSIs for both tasks. To the best of our knowledge, this is the first attempt to classify diffuse gliomas via modeling the interaction of histology and molecular markers.

Our main contributions are: (1) We propose a multi-task multi-instance learning framework to jointly predict molecular markers and histology and finally classify diffuse glioma, reflecting the new paradigm of pathology diagnosis. (2) We design a CPLC-Graph network to model the intra-omic relationship of multiple molecular markers.  
(3) We design a DCC learning strategy to model the inter-omic interaction between histology and molecular markers for glioma classification.

\begin{figure*}[!t]
\centering
\includegraphics[width=1.0\linewidth]{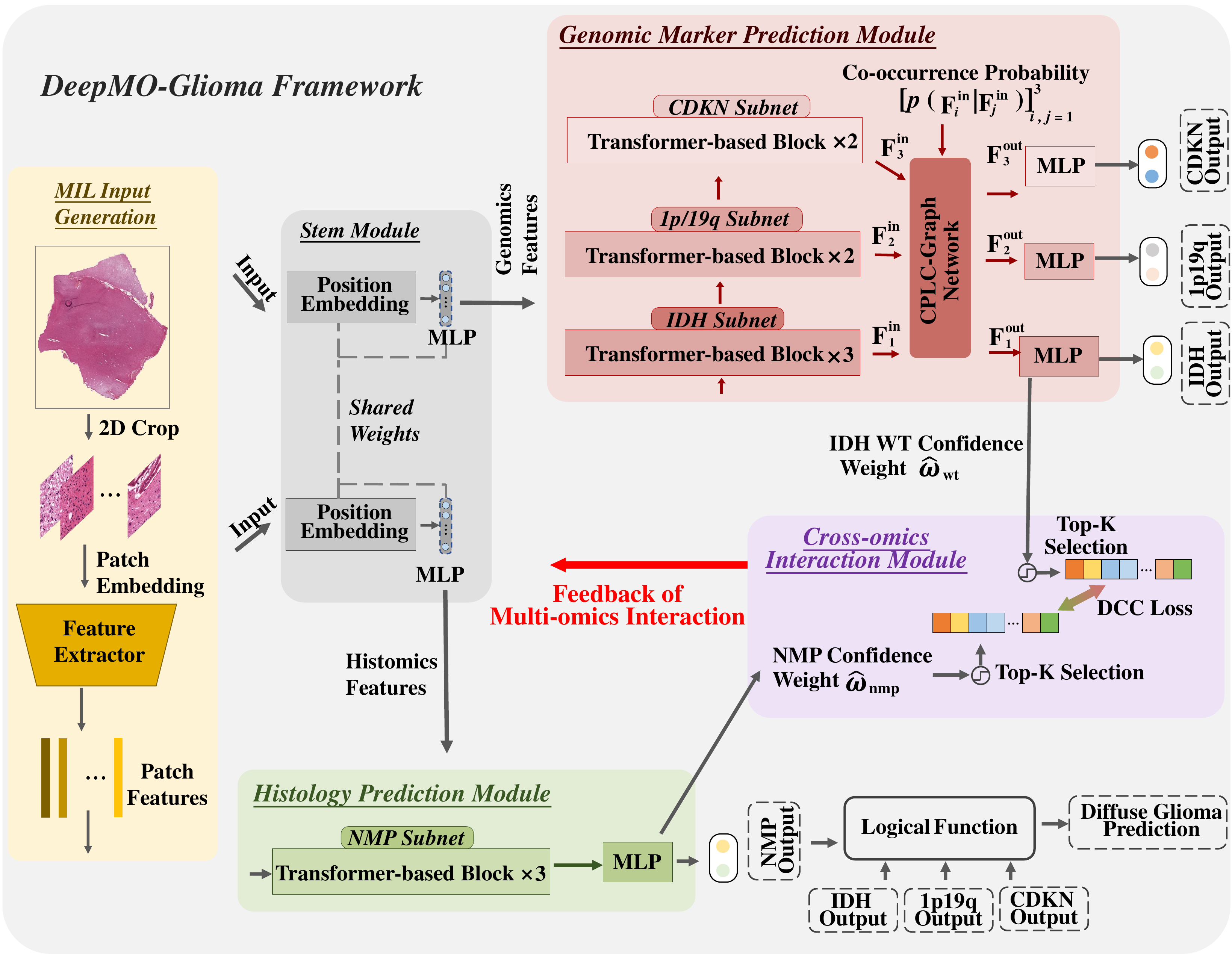}

\caption{\footnotesize Architecture of DeepMO-Glioma.}

\label{fullnet}
\end{figure*}

\section{Preliminaries}

\noindent\textbf{Database:} 
We use publicly available TCGA GBM-LGG dataset \cite{tcga}. Following \cite{lu2021data}, we remove the WSIs of low quality or lack of labels. Totally, we include 2,633 WSIs from 940 cases, randomly split into training (2,087 WSIs of 752 cases), validation (282 WSIs of 94 cases) and test (264 WSIs of 94 cases) sets. All the WSIs are crop into patches of size 224px $\times$ 224px at  0.5 $\mu$m $\textrm{px}^{-1}$.

\noindent\textbf{Training labels:} 
Original lables for genomic markers and histology of WSIs are obtained from TCGA database \cite{tcga}. According to the up-to-date WHO criteria \cite{louis20212021},  we generate the classification labels for each case as grade 4 glioblastoma (defined as IDH widetype), oligodendroglioma (defined as IDH mutant and 1p/19q co-deletion), grade 4 astrocytoma (defined as IDH mutant, 1p/19q non co-deletion with CDKN HOMDEL or NMP), or low-grade astrocytoma (other cases).

\section{Methodology}

Figure \ref{fullnet} illustrates the proposed DeepMO-Glioma. As shown above, the up-to-date WHO  criteria incorporates molecular markers and histology features. Therefore, our model is designed to jointly learn the tasks of predicting molecular markers and histology features in a unified framework. DeepMO-Glioma consists four modules, i.e, stem, genomic marker prediction, histology prediction and cross-omics interaction.  Given the cropped patches $\{\mathbf{X}_{i}\}_1^{N}$  as the input,  DeepMO-Glioma outputs 1) the status of molecular markers, including IDH mutation $\hat{l}_{idh} \in \mathbb{R}^{2} $, 1p/19q co-deletion $\hat{l}_{1p/19q} \in \mathbb{R}^{2} $ and CDKN HOMDEL $\hat{l}_{cdkn} \in \mathbb{R}^{2}$, 2) existence of NMP $\hat{l}_{nmp} \in \mathbb{R}^{2} $ and 3) final diagnosis of diffuse gliomas $\hat{l}_{glio} \in \mathbb{R}^{4}$.

\subsection{Hierarchical multi-task multi-instance learning} To extract global information from input  $\{\mathbf{X}_{i}\}_1^{N}$, we propose a hierarchical multi-task multi-instance learning (HMT-MIL)  framework for both histology and molecular marker predictions. Different from methods using one \cite{zhang2022mutual} or several \cite{chen2020pathomic,xing2022discrepancy} representative patches per slide, HMT-MIL framework can extract information from N=2,500 patches per WSI via utilizing the MIL  learning paradigm with transformer blocks \cite{dosovitskiy2020image} embedded. Note for WSIs with patch number< N, we adopt a biological repeat strategy for dimension alignment.

\begin{figure*}[!t]
\centering
\includegraphics[width=1.0\linewidth]{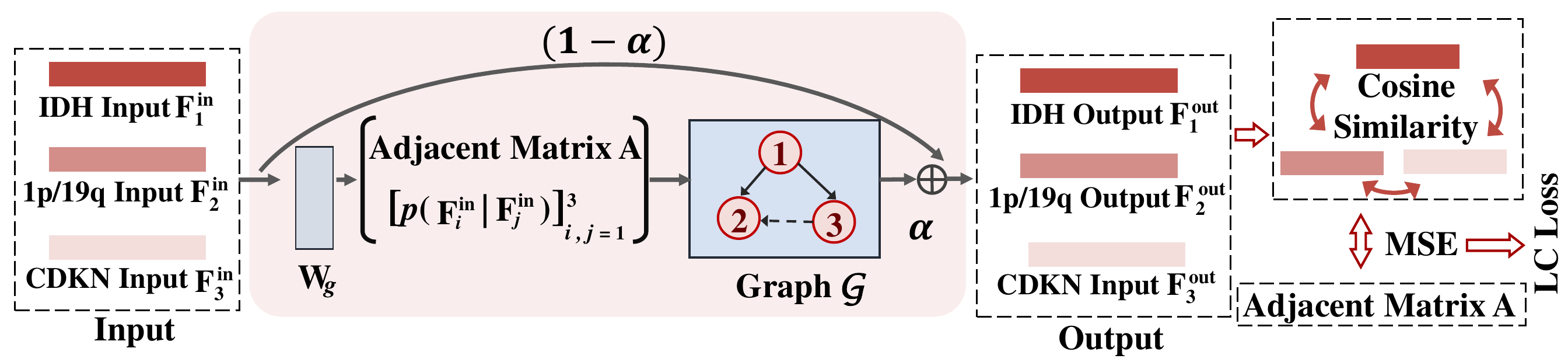}
\caption{\footnotesize Pipelines of CPLC-Graph network (a) and DCC loss (b).}

\label{fullnet2}
\end{figure*}

\subsection{Co-occurrence probability-based, label-correlation graph}

In predicting molecular markers, i.e., IDH, 1p/19q and CDKN,  existing MLC methods based on label correlation may ignore the co-occurrence of the labels. We proposed a co-occurrence probability-based, label-correlation graph (CPLC-Graph) network and a label correlation (LC) loss for intra-omic modeling of the co-occurrence probability of the three markers.
   
\noindent\textbf{1) CPLC-Graph network:} CPLC-Graph (Figure \ref{fullnet2}) is defined as $ \mathcal{G}=(\mathbf{V},\mathbf{E})$, where  $\mathbf{V}$  indicates the nodes, while $\mathbf{E}$ represents the edges. Given the intermediate features in predicting the three molecular markers subnets  $\mathbf{F}^{\rm in}=[\mathbf{F}^{\rm in}_{i}]_{i=1}^{3} \in \mathbb{R}^{3\times C}$ as input nodes, we construct a co-occurrence probability based correlation matrix $\mathbf{A} \in \mathbb{R}^{3\times 3}$ to reflect the relationships among each node feature, with a weight matrix $\mathbf{W}_{g} \in \mathbb{R}^{C\times C}$ to update the value of $\mathbf{F}^{\rm in}$. Formally, the output nodes $\mathbf{F}^{\rm mid} \in \mathbb{R}^{3\times C}$ are formulated by a single graph convolutional network layer as

\begin{equation}
\label{E1}
\mathbf{F}^{\rm mid} = \mathcal{\delta} (\mathbf{A} \mathbf{F}^{\rm in} \mathbf{W}_{g}) , \text{where} \mathbf{A} = [A_{i}^{j}]_{i,j=1}^{3}, A_{i}^{j} = \frac{1}{2}\big (p (\mathbf{F}^{\rm in}_{i}|\mathbf{F}^{\rm in}_{j})+p (\mathbf{F}^{\rm in}_{j}|\mathbf{F}^{\rm in}_{i})\big).
\end{equation}
In \eqref{E1}, $\mathcal{\delta} (\cdot)$ is an activation function and $p (\mathbf{F}^{\rm in}_{i}|\mathbf{F}^{\rm in}_{j})$ denote the probability of the status of $i$-th marker given the status of $j$-th marker. Besides, residual structure is utilized to generate the final output $\mathbf{F}^{\rm out}$ of CPLC-Graph network, defined as  
$\mathbf{F}^{\rm out} = \alpha \mathbf{F}^{\rm mid} + (1-\alpha) \mathbf{F}^{\rm in}, $
where $\alpha$ is a graph balancing hyper-parameter.

\noindent\textbf{2) LC loss:} In order to fully exploit the co-occurrence probability of different molecular markers, we further devise the LC loss that constrains the similarity between any two output molecular markers $\mathbf{F}^{\rm out}_{i}$ and $\mathbf{F}^{\rm out}_{j}$ to approach their correspondent co-occurrence probability  $A_{i}^{j}$. 
 Formally, the LC loss is defined as
\begin{equation}
\label{loss0}
\mathcal{L}_{\rm LC}= \mathcal{MSE} (\mathbf{A},\mathbf{D}_{\rm cos}), \text{where} \mathbf{D}_{\rm cos} =[D_{cos}^{i,j}]_{i,j=1}^{3}, D_{cos}^{i,j}= \frac{(\mathbf{F}^{\rm out}_{i})^\top \mathbf{F}^{\rm out}_{j}} { \left\| \mathbf{F}^{\rm out}_{i} \right\| \left\|\mathbf{F}^{\rm out}_{j} \right\| }. 
\end{equation}
In \eqref{loss0}, $\mathcal{MSE}$ denotes the function of mean square error, while $D_{cos}^{i,j}$ is the cosine similarity of features $\mathbf{F}^{\rm out}_{i}$ and $\mathbf{F}^{\rm out}_{j}$ .

\subsection{Dynamical confidence constraint}
We design a dynamical confidence constraint (DCC)  strategy to model the interaction between molecular markers and histological features. Taking IDH and NMP as an example,
the final outputs for IDH widetype\footnote{Note that, IDH widetype is incorporated in diagnosing glioblasoma in current clinical paradigm; while previously, diagnosis of glioblasoma is puley based on NMP} and NMP predictions can be defined as $\hat{l}_{wt} = \sum\nolimits_{n=1}^{N} \omega_{wt}^{n} f_{wt}^{n}$ and  $\hat{l}_{nmp} = \sum\nolimits_{n=1}^{N} \omega_{nmp}^{n} f_{nmp}^{n}$, respectively. Note that $f_{wt}^{n}$ and $\omega_{wt}^{n}$ are values of the extracted feature and the corresponding decision  weight of $n$-th patch, respectively. We then reorder $[\omega_{wt}^{n}]_{n=1}^{N}$ to $[\hat{\omega}_{wt}^{n}]_{n=1}^{N}$ based on their values. Similarly,  we obtain $[\hat{\omega}_{nmp}^{n}]_{n=1}^{N}$ for NMP confidence weights.

Based on ordered confidence weights, we constrain the prediction networks of histology and molecular markers to focus on the WSI areas important for both predictions, thus modeling inter-omic interactions. Specifically, we achieve the confidence constraint through a novel DCC loss focusing on top $K$ important patches for both prediction. Formally, the DCC loss in $m$-th training epoch is defined as:
\begin{equation}
\label{loss1}
\mathcal{L}_{\rm DCC}= \frac{1}{2K_m} \sum\nolimits_{k=1}^{K_m} \big( \mathcal{S}(\hat{\omega}_{wt}^{k},\hat{\omega}_{nmp})+\mathcal{S}(\hat{\omega}_{nmp}^{k},\hat{\omega}_{wt})\big),
\end{equation} 
where $\mathcal{S}(\hat{\omega}_{wt}^{k},\hat{\omega}_{nmp})$ is the indicator function taking the value 1 when the $k$-th important patch of IDH widetype is in the set of top $K_m$ important patches for NMP, and vice versa. In addition, to facilitate the learning process with DCC loss, we adopt a curriculum-learning based training strategy dynamically focusing on hard-to-learn patches,  regarded as the patches with higher decision importance weight, as patches with lower confidence weight, e.g., patches with fewer nuclei, are usually easier to learn in both tasks. Hence, $K_m$ is further defined as 

\begin{equation}
\label{eq2}
K_m=K_0 \beta^{\lfloor\frac{m}{m_0} \rfloor }.
\end{equation} 
In \eqref{eq2}, $K_0$ and $m_0$ are hyper-parameters to adjust  $\mathcal{L}_{\rm DCC}$ in training process.

\section{Experiments \& Results}

\subsection{Implementation details}
The proposed DeepMO-Glioma is trained on the training set for 70 epochs, with batch size 8 and learning rate 0.003 with Adam optimizer~\cite{kingma2014adam} together with the
weight decay. Key hyper-parameters are in Table I of supplementary material. All hyper-parameters are tuned to achieve the best performance over the validation set. All experiments are conducted on a computer with an Intel(R) Xeon(R) E5-2698 CPU @2.20GHz, 256GB RAM and 4 Nvidia Tesla V100 GPUs. Additionally, our method is implemented on PyTorch with the Python environment.

\begin{table}[!t]
\scriptsize
\centering
\caption{Performance of classifying glioma based on WHO 2021 criteria \cite{bale20222021}.}
\label{expe:label}
\begin{tabular}{cccccccccccccccccc}
    \toprule
    \multirow{1}*{Method} & \multicolumn{11}{c}{Diffuse glioma classification, \% }   &\multicolumn{1}{p{1em}}{} & \multicolumn{5}{c}{Ablation study (W/O), \% } \\
    \specialrule{0em}{1pt}{1pt}
    \cline{2-12}  
    \cline{14-18}
    \specialrule{0em}{2pt}{2pt}
      & Ours.  &\multicolumn{1}{p{.4em}}{} &CLAM  &\multicolumn{1}{p{0.2em}}{} & TransMIL  &\multicolumn{1}{p{0.2em}}{} & ResNet* &\multicolumn{1}{p{0.2em}}{} & DenseNet* &\multicolumn{1}{p{0.2em}}{} & VGG-16* &\multicolumn{1}{p{0.5em}}{}& Graph &\multicolumn{1}{p{0.2em}}{}& LC loss &\multicolumn{1}{p{.2em}}{} & DCC\\

    \midrule

Acc. & \textbf{77.3} &\multicolumn{1}{p{.4em}}{} & 71.2 &\multicolumn{1}{p{0.2em}}{} & 68.2 &\multicolumn{1}{p{0.2em}}{}& 59.1 &\multicolumn{1}{p{0.2em}}{}& 62.9 &\multicolumn{1}{p{0.2em}}{}& 60.6 &\multicolumn{1}{p{0.1em}}{} & 65.5 &\multicolumn{1}{p{0.5em}}{}& 71.2 &\multicolumn{1}{p{1em}}{} & 68.2 \\

Sen. & \textbf{76.0} &\multicolumn{1}{p{.4em}}{}& 62.9 &\multicolumn{1}{p{0.2em}}{} & 60.2 &\multicolumn{1}{p{0.2em}}{} & 51.6 &\multicolumn{1}{p{0.2em}}{} & 52.5 &\multicolumn{1}{p{0.2em}}{} & 49.8 &\multicolumn{1}{p{0.1em}}{} & 47.7&\multicolumn{1}{p{0.5em}}{} & 61.0 &\multicolumn{1}{p{1em}}{}& 59.8 \\

Spec. & \textbf{86.6} &\multicolumn{1}{p{.4em}}{}& 82.9 &\multicolumn{1}{p{0.2em}}{} & 79.9 &\multicolumn{1}{p{0.2em}}{} & 71.4 &\multicolumn{1}{p{0.2em}}{} & 83.5 &\multicolumn{1}{p{0.2em}}{}& 74.5 &\multicolumn{1}{p{0.1em}}{} & 83.0 &\multicolumn{1}{p{0.5em}}{} & 82.3 &\multicolumn{1}{p{1em}}{}& 84.7 \\

$\mathrm{F_{1}}$-score & \textbf{71.0} &\multicolumn{1}{p{.4em}}{}& 60.0 &\multicolumn{1}{p{0.2em}}{} & 59.2 &\multicolumn{1}{p{0.2em}}{} & 51.6 &\multicolumn{1}{p{0.2em}}{} & 50.9 &\multicolumn{1}{p{0.2em}}{}& 49.6 &\multicolumn{1}{p{0.1em}}{} & 49.4 &\multicolumn{1}{p{0.5em}}{} & 61.2 &\multicolumn{1}{p{1em}}{}& 53.5 \\

\bottomrule

\multicolumn{18}{l}{{* In this paper, all these methods are slighted modified to adjust to the MIL setting.}}\\

\end{tabular}

\end{table}

\subsection{Performance evaluation}

\noindent\textbf{1) Glioma classification.} We compare our model with five other state-of-the-art methods: CLAM\cite{lu2021data}, TransMIL \cite{shao2021transmil},  ResNet-18 \cite{he2016deep},  DenseNet-121 \cite{huang2017densely} and  VGG-16 \cite{simonyan2014very}. Note CLAM  \cite{lu2021data} and TransMIL \cite{shao2021transmil} are MIL framework, while others are commonly-used image classification methods, set as our baseline. 
The left panel of Table \ref{expe:label} shows that DeepMO-Glioma performs the best, achieving at least 6.1\%, 13.1\%, \ 3.1\% and 11.0\% improvement over other models in accuracy, sensitivity, specificity and AUC, respectively,  indicating that our model could effectively integrate molecular markers and histology in classifying diffuse gliomas.

\begin{figure*}[!t]
\centering
\includegraphics[width=.7\linewidth]{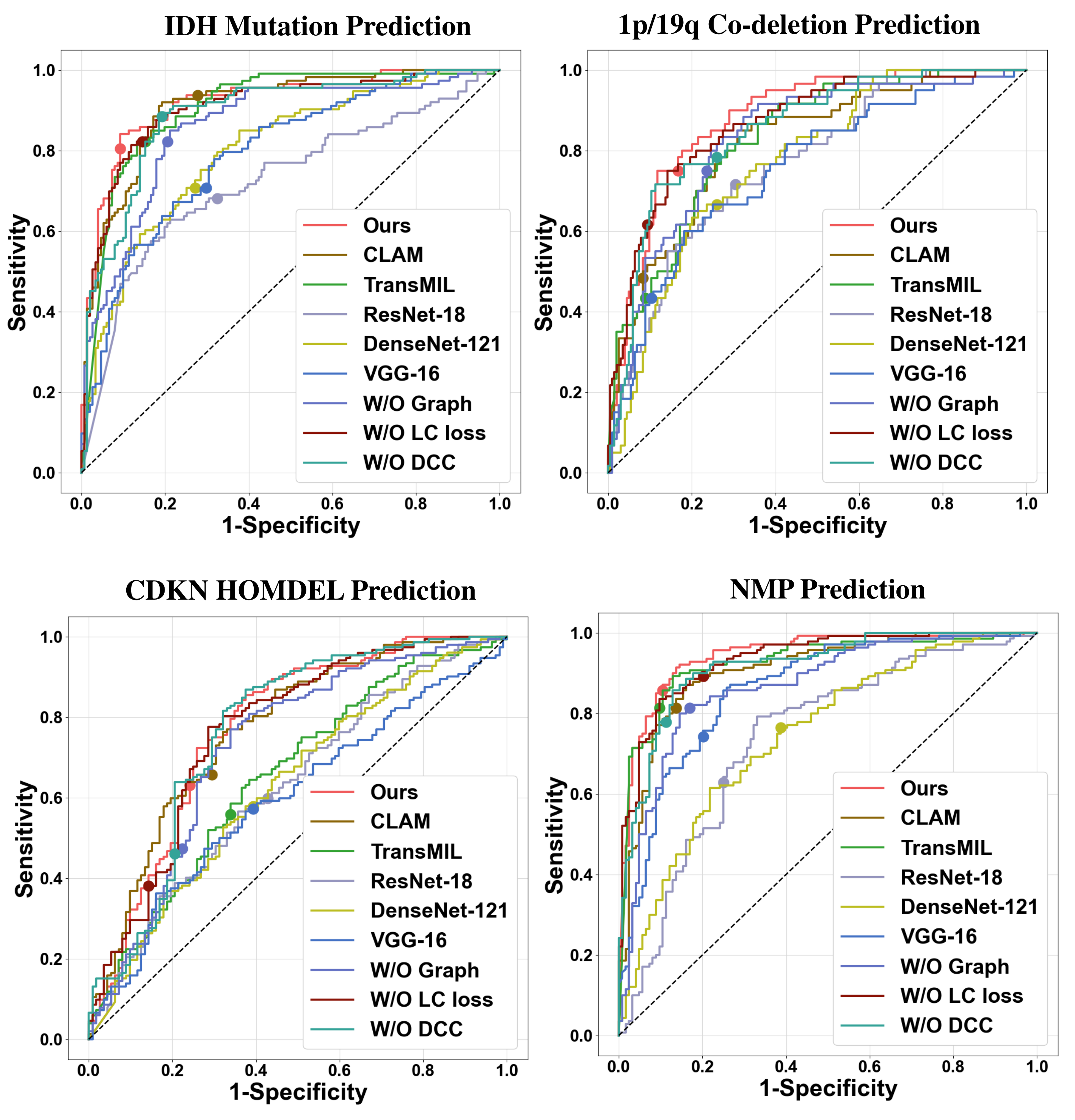}

\caption{
ROC curves of our model, comparison and ablation models for predicting IDH, 1p/19q, CDKN and NMP. 
}

\label{expe:fig}
\end{figure*}

\noindent\textbf{2) Predictions of genomic markers and histology features.}
From the left panel of Table \ref{tab:addlabe2}, we observe that DeepMO-Glioma achieves the AUC of 92.0\%, 88.1\%, 77.2\% and 94.5\% for IDH mutation, 1p/19q co-deletion, CDKN HOMDEL and NMP prediction, respectively, considerably better than all the comparison models. Figure \ref{expe:fig} (b) plots the ROC curves of all models. demonstrating the superior performance of our model over other comparison models.

\noindent\textbf{3) Network interpretability.} An additional visualization experiment is conducted based on patch decision scores to test the interpretability of our method. Due to the page limit, the results are presented in supplementary Figure 1.

\begin{table*}[!t]
\scriptsize
  \centering
  \caption{Performance in predicting genomic markers,  histology and ablation studies. 
}

    \begin{tabular}{|l|l|cccccc|ccc|}
 \toprule
     &  & Ours. & CLAM & TransMIL  & ResNet & DenseNet & VGG-16 & No Graph & No LC loss & No DCC \\

	\cline{2-11}	
    & \quad Acc.  &   \textbf{86.4}  &   81.4 & 83.7  &   67.8 &   72.0  & 70.5 &  80.7 & 84.1 &  84.1 \\
    & \quad Sen.  &   80.5  &   93.8 & 82.3  &   68.1 &   70.8  & 70.8 &  82.3 & 82.3 &  88.5    \\
    & \quad Spec. &   90.7  &   72.2 & 84.8  &   67.5 &   72.8  & 70.2 &  79.5 & 85.4 &  80.8  \\
     
     \multirow{-4}{*}{ \rotatebox{90}{ \textbf{IDH}}}  & \quad  AUC &   \textbf{92.0}  &   91.1 & 90.7  &   72.7 &   80.3  & 79.6 &  86.1  & 90.8 &  89.1  \\ 
    \toprule
     & \quad Acc.  &   81.4  &   81.8 & 80.3  &   70.1 &   72.3  & 79.2 &  76.1 & \textbf{84.1} &  75.0 \\
     & \quad Sen.  &   75.0  &   48.3 & 43.3  &   71.7 &   66.7  & 43.3 &  75.0 & 61.7 &  78.3   \\
     & \quad Spec. &   83.3  &   91.7 & 91.2  &   69.6 &   74.0  & 89.7 &  76.5 & 90.7 &  74.0  \\
     
     \multirow{-4}{*}{ \rotatebox{90}{ \textbf{1p19q}}}  & \quad  AUC &   \textbf{88.1}  &   82.0 & 82.9  &   76.8 &   77.1  & 75.5 &  83.0  & 86.7 &  85.2  \\
   \toprule

     & \quad Acc.  &   \textbf{68.6}  &   67.8 & 60.2  &   58.7 &   59.1 & 58.7 &  60.2 & 58.3 &  60.2 \\
     & \quad Sen.  &   63.2  &   65.8 & 55.9  &   59.9 &   57.9  & 57.2 &  47.4 & 38.2 &  46.1    \\
     & \quad Spec. &   75.9  &   70.5 & 66.1  &   57.1 &   60.7  & 60.7 &  77.7 & 85.7 &  79.5  \\
     
     \multirow{-4}{*}{ \rotatebox{90}{ \textbf{CDKN}}}  & \quad  AUC &   \textbf{77.2}  &   77.0 & 65.5  &   62.8 &   62.9  & 59.9 &  72.6  & 76.7 &  76.7  \\

    \toprule
     & \quad Acc.  &   \textbf{87.5}  &   83.7 & 85.6  &   68.6 &   69.3  & 76.9 &  82.2 & 84.8 &  83.0 \\
     & \quad Sen.  &   85.7  &   81.4 & 81.4  &   62.9 &   76.4  & 74.3 &  81.4 & 89.3 &  77.9    \\
     & \quad Spec. &   89.5  &   86.3 & 90.3  &   75.0 &   61.3  & 79.8 &  83.1 & 79.8 &  88.7  \\
     
     \multirow{-4}{*}{ \rotatebox{90}{ \textbf{NMP}}}  & \quad  AUC &   \textbf{94.5}  &   90.7 & 92.7  &   74.0 &   74.7  & 86.1 &  86.7  & 93.4 &  91.7 \\

    \toprule
    \end{tabular}%
  \label{tab:addlabe2} 
  
\end{table*}

\subsection{Results of ablation experiments}

\noindent\textbf{1) CPLC-Graph network.} 
The right panels of Table \ref{expe:label} shows that, by setting graph balancing weight $\alpha$ to 0 for the proposed CPLC-Graph, the accuracy, sensitivity, specificity and $\mathrm{F_{1}}$-score decreases by 7.8\%, 29.0\%, 3.6\% and 21.6\%, respectively. Similar results are observed for the prediction tasks of molecular markers and histology (Table \ref{tab:addlabe2}). Also, the ROC curve of removing the CPLC-Graph network is shown in Figure \ref{expe:fig}. These results indicate the utility of the proposed CPLC-Graph network.

\noindent\textbf{2) LC loss.} 
The right panels of Table \ref{expe:label} shows that the performance after removing LC loss decreases in all metrics, causing a reduction of 6.1\%, 15.0\%, 4.3\% and 9.8\%, in accuracy, sensitivity, specificity and $\mathrm{F_{1}}$-score, respectively. Similar results for the tasks of molecular marker and histology prediction are observed in the right panel of Table \ref{tab:addlabe2} with ROC curves in Figure \ref{expe:fig}, indicating the effectiveness of the proposed LC loss.

\noindent\textbf{3) DCC loss.}  From Table \ref{expe:label}, we observe that the proposed DCC loss improves the performance in terms of accuracy by 9.1\%. Similar results can be found for sensitivity, specificity and $\mathrm{F_{1}}$-score. From Table \ref{tab:addlabe2}, we observe that the AUC decreases 2.9\%, 2.9\%, 0.5\% and 2.8\% for the prediction of IDH, 1p/19q, CDKN and NMP, respectively, when removing the DCC loss. Such performance is also found in comparing the ROC curves in Figure \ref{expe:fig}, suggesting the importance of the DCC loss for all the tasks.

\section{Summary }
 
The paradigm of pathology diagnosis has shifted to integrating molecular makers with histology features. In this paper, we aim to classify diffuse gliomas under up-to-date diagnosis criteria, via jointly learning the tasks of molecular marker prediction and histology classification.
Inputting histology WSIs, our model incorporates a novel HMT-MIL framework to extract global information for both predicting both molecular markers and histology. We also design a CPLC-Graph network and a DCC loss to model both intra-omic  and inter-omic interactions. Our experiments demonstrate that our model has achieved superior performance over other state-of-the-art methods, serving as a potentially useful tool for digital pathology based on WSIs in the era of molecular pathology.

\clearpage

\bibliographystyle{splncs04}
\bibliography{ref}

\clearpage

\vspace{5em}
\centerline{\LARGE Supplementary Materials}

\begin{figure}[h]
\centering
\includegraphics[width=0.99\linewidth]{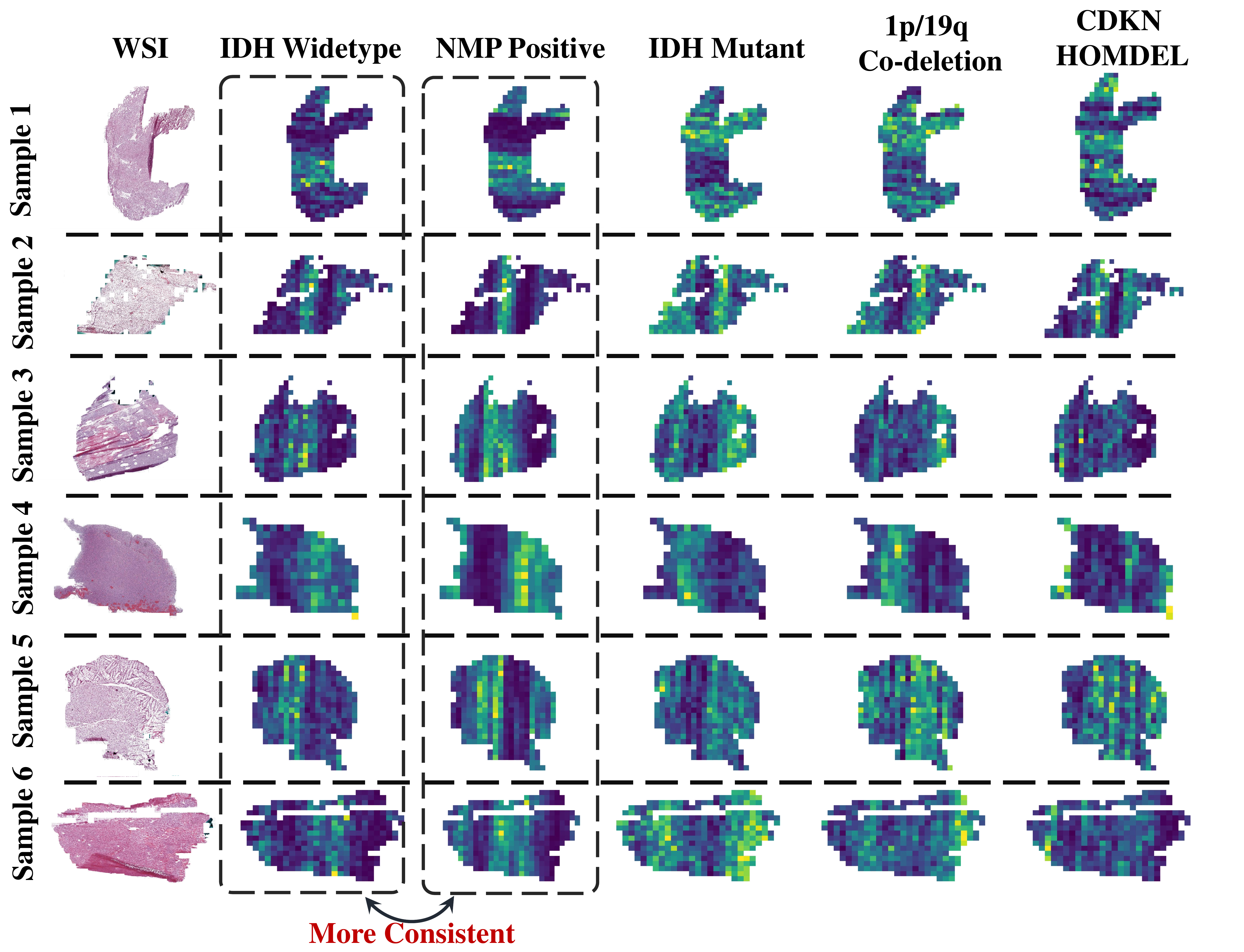}
\caption{Visualization maps of DeepMO-Glioma predicting molecular markers and histology. Based on patch decision scores of correctly predicted WSIs from IDH widetype glioblastoma,  it is observed  that our method generates more consistent maps for
predicting IDH widetype (molecular marker) with NMP positive (histology feature), compared to other molecular markers. This finding is consistent with the up-to-date  diagnostic criteria where glioblastoma is predominately IDH wildtype and NMP positive in histology, suggesting our success in modeling inter-omic co-occurrence, which thus could indicate the interpretability when integrating molecular markers with histology for clinical diagnosis.
}
\label{database}
\end{figure}

\begin{table}[h]
\caption{Implementation details of our proposed method.}\label{Parameters}
\begin{center}
\begin{tabular}{| l|l|}
 \hline
 
   Number of features $C$ for the input nodes of CPLC-Graph  & $512$ \\
   Graph balancing weight $\alpha$ &  $0.1$ \\
   $\beta$ in adjusting  $\mathcal{L}_{\rm DCC}$ in equation (4) & $0.85$ \\
   $K_0$ in adjusting  $\mathcal{L}_{\rm DCC}$ in equation (4) & $1250$ \\
   $m_0$ in adjusting  $\mathcal{L}_{\rm DCC}$ in equation (4) & $10$ \\
   Exponential decay rate $\beta_{1}$ and $\beta_{2}$ for Adam optimization &  0.9 and 0.999 \\
   Epsilon $\epsilon$ for Adam optimization   &  $1\times10^{-8}$\\
   Weight decay  for Adam optimization   &  $1\times10^{-5}$\\
   
   Inference speed &  17 WSIs per second \\

  \hline

\end{tabular}
\end{center}
\end{table}

\end{document}